# Assessing the impact of tourist attractions through the integration of causal inference and demand-side economic analysis

## A case study of the Sensoria experience museum in Holzminden, Germany

Thomas Wieland
Independent researcher, Freiburg
ORCID: https://orcid.org/0000-0001-5168-9846

## Abstract

Assessing the economic impact of tourist attractions is a central field of research in tourism geography, tourism economics, and regional economics. This kind of research typically adopts a demand-side economic approach, which is frequently based on visitor surveys that ask, for example, about visitors' on-site spending and the nature of their visiting modalities. This approach has several limitations, particularly that it is not based on causal analysis and that the results of surveys could be heavily biased. This study proposes an integrated framework combining causal inference and demand-side economic analysis. As a case study, the impact of the *Sensoria* experience museum in Holzminden, Germany (opened in September 2024) on tourism demand (measured in monthly overnight stays) is examined. Two difference-in-differences (DiD) approaches are employed to quantify the number of additional overnight stays in the treatment city: a conventional DiD model using a control group and a DiD model using a synthetic control unit. The results are converted into industry-specific expenditures, from which the direct and indirect effects of *Sensoria* are determined. The results are mixed: the DiD model detects a significantly positive impact in the first year of operation of the new tourist attraction, whereas the DiD model with the synthetic control unit shows a positive but insignificant treatment effect. The significant DiD estimate corresponds to 4,964 additional overnight stays in the first year. When this is offset against the average expenditure of overnight guests, the result is an additional gross turnover of approximately 0.60 million EUR across the hospitality and retail industries and other services. The resulting direct effects and indirect effects amount to approximately 0.24 and 0.22 million EUR, respectively. However, long-term effects cannot (yet) be determined. This study demonstrates that combining the two approaches mentioned holds promise, yet requires a more in-depth analysis, for which suggestions are also discussed regarding how it could be conducted.

## Author contact

Dr. Thomas Wieland
geowieland@googlemail.com

## Declarations

This research received no external funding. The author declares no conflict of interest.

The author would like to thank Martina and Ernst Wieland for their help in compiling information on the Sensoria case. The corresponding Python analysis script, including the data, is available in the following public GitHub repository: https://github.com/geowieland/sensoria. The analysis may be updated when new data becomes available.

# 1 Introduction

The study of the macroeconomic and regional economic contributions of tourism is a topic of great interest, not least due to the development of tourism satellite accounts (TSA) by public institutions (OECD et al., 2001). However, assessing the economic impact of tourist attractions is a central field of research in tourism geography, tourism economics, and regional economics as well. This especially applies to examining whether public investments have a valuable return on investment, for example, with respect to the hospitality and retail industries, and other sectors. The underlying assumption is that physical or intangible tourist attractions increase the number of day visitors and/or overnight guests, and that their additional spending (e.g., accommodation, shopping) also generates additional regional income. This may refer to the economic effects of specific facilities such as museums or concert halls (e.g., Eckhardt et al., 2026; Llop & Arauzo-Carod, 2012; Plaza, 2006), tourist events (e.g., Cabras et al., 2020; Džupka & Šebová, 2016; Wallimann and Mehr, 2026) or to a particular intangible status of a region, such as its status as a national park, biosphere reserve or World Heritage Site (e.g., Mayer et al., 2010; Majewski et al., 2025; Wang et al., 2024).

Research on the economic impact of tourist attractions typically adopts a demand-side economic approach based on input-output analysis, similar to its application in tourism satellite accounts (OECD et al., 2001). The key idea behind this methodology is to estimate the impact of visitor spending with respect to accommodation, restaurants, shopping, etc. These expenditures generated by a tourist attraction are then decomposed into regional income that is directly generated by visitor spending (*direct effects*, e.g., wages of employees in hotels and restaurants), and the demand for intermediate goods (*indirect effects*, e.g., deliveries from food retailers to restaurants). Tourist spending is usually captured through visitor surveys, and the results are extrapolated to the total number of visitors. Particularly at free-entry events and facilities, surveys are used to identify visitors who are present at the destination specifically *because* of the tourist attraction, thereby distinguishing them from tourists whose visit is not causally linked to the facility being studied (e.g., Cabras et al., 2020; Džupka and Šebová, 2016; Llop and Arauzo-Carod, 2012; Majewski et al., 2025; Mayer et al., 2010).

This approach to assessing the economic impacts of tourist attractions has several limitations, which are also addressed in the relevant literature (e.g., Chang et al., 2015). A key factor here is that no genuine causal-analytical approach is employed, unlike the methods used to evaluate interventions in the context of natural experiments in economic policy and marketing (Goldfarb et al., 2022; Goodman-Bacon, 2021; Wooldridge, 2012). Furthermore, the results depend on the representativeness of the survey, and it is impossible to filter out temporal coincidences with other tourist-related effects (e.g., those resulting from other tourist attractions). There are isolated studies that employ a causal-analytical approach to evaluate tourist attractions but do not estimate demand-side effects (e.g., Wallimann and Mehr, 2026; Wang et al., 2024). This study addresses the issues mentioned above and, using a case study, employs a methodological workflow that combines a causal inference model and a demand-side economic analysis, with the latter being a methodology frequently utilized in assessing the impact of tourist attractions (Chang et al., 2015; Džupka and Šebová, 2016; Mayer et al., 2010).

The case study addresses the impact of a built tourist establishment embedded in the context of a place-branding strategy: The city of Holzminden (approx. 20,000 inhabitants) in the state of Lower Saxony, Germany, operates a city marketing strategy with the slogan "*Stadt der Düfte und Aromen*" ("City of fragrances and flavors"). This branding aims to highlight the fact that the city has been a center of the fragrance and flavorings industry since the mid-19th century; the local economic structure is strongly influenced by a company located there (*Symrise*, approx. 12,700 employees in 2024), which is one of the world's leading manufacturers in this field. In this context, the interactive fragrance and flavor museum *Sensoria* was opened in September 2024. The exhibition space covers 614 $m^2$ on three floors, plus 500 $m^2$ of outdoor space. The experience museum provides information about the history of the development of fragrances and flavorings and the resulting consumer products. Visitors may also test fragrances and mix their own perfume. According to press reports, the city of Holzminden invested 9 million EUR plus 2.2 million EUR in subsidies from the European Union's ERDF (European Regional Development Fund) fund; the annual operating costs are estimated at 0.5 million euros per year (TAH, 2024a). According to official figures, approx. 23,000 visitors were recorded in the first 12 months; in the calendar year 2025, 21,256 visitors were counted (Stadt Holzminden, 2026; TAH, 2025).

It has been theoretically substantiated and empirically demonstrated that the opening of a new tourist attraction increases the all-over attractiveness of a destination, which may lead to an increase in tourist demand. On the one hand, this is due to the effective visitors of the new tourist facility. On the other hand, certain unique tangible or intangible attractions of a city or region may also sharpen its image in the competition among tourist destinations (Camacho-Murillo et al., 2021; Reinhold et al., 2023; Vinyals-Mirabent, 2019). Museums are among the man-made tourist attractions; they are considered an important component of the attractiveness of a tourist destination (Badurina et al., 2025; Camacho-Murillo et al., 2021; Marrocu and Paci, 2013). A well-known example where the opening of a museum substantially strengthened the tourist attraction of a region is the opening of the Guggenheim Museum in Bilbao, Basque Country, Spain, in 1997. This modern art museum, housed in an extraordinary building designed by star architect Frank Gehry, generates an average of approximately 1 million visitors annually and, according to impact analyses, has made a crucial contribution to the prosperity of the originally economically deprived city of Bilbao. This successful development gave rise to the term “Bilbao effect”, which refers to the economic growth of a city or region through a new (built) attraction (Plaza, 2006).

In order to evaluate the economic impact of *Sensoria*, first, an econometric analysis is performed with a difference-in-differences approach, with monthly tourist overnight stays serving as the outcome variable. In this process, various model variants are compared, while other tourism-related effects resulting from events are included as control variables. Second, the direct and indirect effects of the related additional tourist demand are determined; these effects are calculated based on assumed tourist expenditure for the hospitality and retail industries as well as other services.

The following section provides a brief overview of literature towards determining the economic impact of tourist attractions. Section 3 explains the methodology of the current study and the related data sources. Section 4 presents the empirical results and related discussions. The conclusions and limitations are discussed in section 5.

# 2 Literature review and assessment

## 2.1 Assessing the impact of tourist attractions

Assessing the economic impact of tourist attractions typically utilizes a demand-side economic approach, resulting in direct and indirect effects (see the “Introduction” section). Regarding built tourist attractions, for example, Llop & Arauzo-Carod (2012) investigate the economic impact of the *Gaudí Centre* in Reus, Catalonia, Spain, which is a biographical museum, opened in 2007. They distinguish regional income effects during the construction phase and after opening (exploitation phase), while considering both direct and indirect effects with respect to payments and investments by the museum and expenditures by tourists attracted by the museum. Tourist expenditures are calculated based on a previous survey. Assuming a time period of 20 years, they estimate an additional local income effect of between 26.9 and 34.6 million EUR. Scherer and Zwicker-Schwarm (2024) investigate the economic impact of the *Europa Park*, a theme park in Rust, Germany. They take into account orders placed with regional companies, the demand from Europa-Park employees, and the spending of tourists in the region. They calculate that this theme park generates 2.48 million overnight stays outside of its own hotels and that tourist spending amounts to 203 million EUR. Together with the 125 million EUR in intermediate goods incurred by *Europa Park*, this results in 328 million EUR; the park's total revenue is estimated at 647 million EUR. Converted to full-time equivalent jobs (FTE), *Europa Park* thus generates 6,600 jobs in Baden-Württemberg and France, outside the park itself.

The economic impacts of events are investigated in a similar way: For example, Džupka and Šebová (2016) investigate the economic impact of the *White Night Festival* in Košice, Slovakia. The primary data used here are two visitor surveys, while the total number of visitors is estimated using aerial photographs. To generate a kind of counterfactual – i.e., the visitor volume *without* the event under investigation – visitors are asked how important attending the festival is to their stay in Košice, to later weight the expenditures accordingly. Based on this, they determine an impact of 0.17 million EUR (2012) and 0.27 million EUR (2013), whereby direct and indirect effects are added together. In order to measure the economic impact of a local craft beer festival with about 10,000 visitors in 2017 (*Knavesmire Beer Festival*, York, United Kingdom), Cabras et al. (2020) conduct a survey of festival visitors aiming at the expenditure inside and outside the festival, while distinguishing between local and non-local visitors. Finally, they calculate industry-specific direct and indirect effects to estimate the gross value added of the festival. They estimate total expenditures equal to 0.72 million pounds, with 0.41 million pounds spent at the festival, and the rest is attributed to hotels and gastronomy. According to calculations, this expenditure corresponds to a gross value added of between 0.46 and 0.56 million pounds and generates between 14.8 and 17.6 FTE jobs.

Similar approaches are applied with respect to the impact of intangible tourist attractions: Regarding national park tourism in Germany, in a series of studies, the regional economic effects of the related tourist activities are estimated using a standardized methodology towards surveying overnight guests and day visitors (Job et al., 2023; Job et al., 2016; Mayer et al., 2010, and others). For example, Job et al. (2023) found that the *Wadden Sea National Park* recorded 21.7 million visitor days in the 12 months studied, of which 90.7% were overnight guests and 9.3% were day visitors. Based on their expenditure, this resulted in a tourism-related added value of 846.7 million EUR, which corresponds to a total of 34,126 FTE jobs, when summing up direct and indirect effects. In the survey, "national park visitors in the narrower sense" are identified, i.e., those for whom visiting the national park plays a decisive role in choosing their holiday destination; their share of the identified FTE amounts to 4,998 FTE. In a similar way, the same working group evaluates the impact of German biosphere reserves: According to the empirical study, the 18 biosphere reserves in Germany recorded 71.6 million visitor days, which induce 1.97 billion EUR of regional income, with 11% of the visitor days belonging to visitors with a "high affinity" towards biosphere reserves (Majewski et al., 2025).

However, there are also isolated studies that employ a different methodological approach to examine the economic impacts of tourist attractions. With respect to built tourist attractions, a well-known example of impact assessment is the evaluation of the opening of the *Guggenheim Museum Bilbao*. Plaza (2006) performs a time-series analysis of overnight stays, accommodation capacity (beds offered), and employment in the service sector in the host region, and tests for significant changes in trends. Following this analysis, the opening of this museum, which has attracted an average of approximately 1 million visitors per year in the first eight years after opening, induced an average of 61,742 additional tourist overnight stays per month (or 740,904 overnight stays per year) and, through direct and indirect effects, created 907 new jobs (FTE).

More recently, few studies have utilized a causal inference approach to assess the impact of tourist attractions on local tourist demand. For example, Wallimann and Mehr (2026) use a difference-in-differences approach to estimate the impact of the 2025 *UEFA Women's European Championship* in Switzerland; they compare host cities with non-host cities, with overnight stays being the outcome. When including all relevant cities in the model analysis, they find a positive but non-significant effect of 2,536 additional overnight stays. When only considering the main host cities, they detect a significant uplift equal to 3,419 additional overnight stays. However, tourist expenditures or income effects are not derived from this estimated additional tourist demand. Wang et al. (2024) use a difference-in-differences approach to examine the effect of defining Chinese regions as World Heritage sites (WHS) on tourism income in those regions. They explicitly distinguish between the period when local governments prepare for WHS status (initiation) and the period when WHS status comes into effect (inscription). They use domestic tourism revenue and foreign tourism revenue as outcomes. Interestingly, they find a significant positive impact on domestic tourism revenues in the initiation phase and a negative effect in the inscription phase, while they cannot detect a positive effect on foreign tourism revenues.

Eckhardt et al. (2026) employ a different causal inference approach – namely, the synthetic control method – to examine the impact of Hamburg's *Elbphilharmonie* (Germany), an elaborately constructed and prominent concert hall, on tourism. This approach compares the actual trend of tourist overnight stays in Hamburg with the counterfactual figures – that is, the number of overnight stays that would have occurred had the *Elbphilharmonie* not opened. The authors find that, between January 2017 and June 2025, the opening generated approximately 18 million additional overnight stays.

## 2.2 Identification of limitations and research gaps

In summary, regarding the literature on the economic impacts of tourist attractions, it can be said that 1) the definitive standard procedure for assessment is demand-side economic analysis, 2) tourist volume, visitor arrangements of tourists, and tourist behavior are usually determined through visitor surveys, and 3) the hypothesis of positive effects of tourist attractions on tourist demand are usually confirmed. However, such impact analyses may have substantial limitations, especially when based on visitor surveys (see, e.g., Chang et al., 2015). *First*, while survey studies have the advantage that all types of visitors are reached (e.g., including day visitors), the question arises as to how representative these surveys are. Any kind of bias in a survey may significantly distort the result once visitor shares and expenditures are extrapolated.

*Second*, the question arises whether these studies contain an adequate counterfactual, i.e., whether the tourist demand that *would* exist *without* the investigated tourist attraction can be accurately represented (Goldfarb et al., 2022). Since it is not possible to observe a city or region at the same time with and without the regarded attraction, it is only possible to identify, through surveying, visitors who are there as tourists *because* of the attraction. This is either asked about directly (e.g., Job et al., 2023) or respondents are asked to rate the relevance of the attraction to their travel decision, so that the survey results can subsequently be weighted accordingly (e.g., Džupka and Šebová, 2016), or a broad distinction is made between visitors from outside the area and those from the host region (e.g., Cabras et al., 2020). Regarding the first point mentioned (the representativeness of the survey), it is questionable whether the actual "additional" visitors can be identified in this way. For enclosed tourist attractions with an admission fee, the total, precisely recorded number of visitors is often used as a basis and counted as "additional" visitors to the region (e.g., Scherer and Zwicker-Schwarm, 2024).

However, it cannot be determined with certainty that all recorded visitors would not have been in the host region without the attraction under study.

*Third*, identifying causal effects also involves controlling for potential confounders, i.e., other variables which influence the outcome under study (Li et al.,2024). With respect to tourism, this may involve changes in the supply structure of accommodation providers in the region under study: For example, Jiménez et al. (2022) show that increases in the supply of *Airbnb* properties stimulate local tourism demand. Furthermore, it is possible for the impacts of various tourist attractions to overlap, for example, when a tourism-related opening takes place at the same time as an unrelated event. It is impossible, or extremely difficult, to distinguish these effects from one another using the established methodology described above.

This causality gap is addressed by the few impact studies using a causal inference approach (e.g., Eckhardt et al., 2026; Wang et al., 2024; Wallimann and Mehr, 2026). These approaches include a counterfactual, more precisely: a control group, and allow for controlling confounders by including covariates (Goldfarb et al., 2022; Li et al., 2024; Wooldridge, 2012). However, these studies typically do not provide an estimation of regional income effects in the relevant industries such as hospitality and retailing. A kind of hybrid form of both approaches is found only in Plaza's (2006) study assessing the impact of the *Guggenheim Museum*; however, this study does not employ a counterfactual, but rather determines additional tourist overnight stays and FTE jobs through time-series analysis.

# 3 Own impact assessment approach

The present study combines the two approaches of impact analysis with respect to tourist attractions, as identified in section 2. The first step involves examining whether the museum's opening induced additional tourist demand and, if so, the magnitude of this uplift; a causal inference design is used for this purpose. The second step, based on the additional tourist demand, calculates the resulting additional income in the form of direct and indirect effects; this step builds upon the demand-side studies of the impact of tourist attractions.

## 3.1 Difference-in-differences (DiD) analysis

To test and quantify the shift in tourist demand due to the opening of *Sensoria*, a difference-in-differences approach (DiD) is applied. Building on this, a difference-in-differences analysis using a synthetic control unit (Synthethic DiD) is conducted as an alternative analytical model. These modeling approaches are typically used in the causal analysis of natural experiments, for example, to evaluate policy interventions or corporate marketing measures (Goldfarb et al., 2022; Goodman-Bacon, 2021; Li et al., 2024; Wooldridge, 2012). In both cases, two models are estimated in which the treatment is divided differently over time: 1) for the entire period the *Sensoria* Museum has been open, and 2) separately for the initial opening year and the subsequent period. This approach aims to examine whether there is a specific opening effect or novelty effect that differs from the museum's permanent presence as a tourist attraction. Such an effect was observed, for example, at the *Guggenheim Museum* in Bilbao: in the first full year (1998) after its opening, the museum recorded 1.3 million visitors; subsequently, visitor numbers fell and stabilized at around 900,000 annually (Plaza, 2006).

The outcome representing local tourist demand used here is the monthly tourist overnight stays. This variable is transformed using the natural logarithm, which changes its potentially right-skewed distribution towards a normal distribution and alters the interpretation of the regression coefficients as (approximate) percentage changes in $Y$ (Jiménez et al., 2022; Wang et al., 2026; Wooldridge, 2012). Fixed effects for the observational units and time points are included to address unobserved heterogeneity, i.e., to compensate for unobserved time-invariant differences between the observational units and unobserved temporal effects independent of the unit, respectively (Baker et al., 2022; Goodman-Bacon, 2021; Wooldridge, 2012). Since a tourist demand variable is being modeled here, the fixed effect for the time periods is particularly important to capture seasonal effects in tourism which are relevant for all observational units and are not attributable to tourist attractions (Zvaigzne et al., 2022). In addition, we include control variables that vary over time *and* region (see below).

The difference-in-differences equation for the full-time period is thus:

$$\ln Y_{jt} = \alpha_j + \lambda_t + \delta\, Sensoria_{jt} + \sum_{c=1}^{C} \beta_c\, V_{c_{jt}} + \varepsilon_{jt} \qquad (1)$$

where $Y_{jt}$ equals the dependent variable (overnight stays) in municipality $j$ in month $t$, $Sensoria_{jt}$ is a binary variable which indicates whether the investigated museum is opened in municipality $j$ in month $t$, $V_{c,jt}$ is the $c$-th control variable with respect to municipality $j$ and month $t$, $\alpha_j$ and $\lambda_t$ are fixed effects for the observational units (municipalities) and time points (months), respectively, $\delta$ and $\beta_c$ are regression coefficients to be estimated, with $\delta$ being the *average treatment effect*, and $\varepsilon_{jt}$ is the stochastic disturbance term.

The second model equation, where the time since the opening is divided into two sections, is:

$$\ln Y_{jt} = \alpha_j + \lambda_t + \delta_1\, Sensoria1_{jt} + \delta_2\, Sensoria2_{jt} + \sum_{c=1}^{C} \beta_c\, V_{c_{jt}} + \varepsilon_{jt} \qquad (2)$$

where $Sensoria1_{jt}$ is a binary variable which indicates whether the investigated museum is opened in municipality $j$ in month $t$ in the first opening year (September 2024 to August 2025), $Sensoria2_{jt}$ represents the same for the subsequent period (from September 2025 onwards), and $\delta_1$ and $\delta_2$ are the corresponding average treatment effects.

Since the supply of tourist accommodation also creates tourist demand to a certain extent (see, e.g., Jiménez et al., 2022), the number of bed days offered is integrated as a control variable, also in log-transformed form. An example of why the availability of overnight accommodation should definitely be taken into account as a control variable is that in September 2024 – coinciding with the Sensoria opening but otherwise independent of it – a local hotel in the treatment city was expanded and the number of rooms on offer was doubled from 35 to 70 (TAH, 2024b). Furthermore, many tourism studies show that events may generate additional tourist demand (Cabras et al., 2020; Džupka and Šebová, 2016; Wallimann and Mehr, 2026). Therefore, the *event days* at the municipal level are included in the model, divided into four control variables distinguishing between types of events, namely local and regional events, Christmas markets, and two specific long-term events of national relevance which are hosted in one city of the control group (see section 3.4).

Two types of robustness checks for the regression models are conducted: A key requirement for deriving a causal effect using a difference-in-differences analysis is the *parallel trends assumption*; this means that it must be assumed that the trend in the outcome in the treatment and control groups is not significantly different, which can only be tested for the pre-treatment period (Goldfarb et al., 2022; Wang et al., 2024). To ensure this, an additional regression model for the pre-treatment period is estimated, incorporating the temporal trend of both groups and an interaction between group and time:

$$\ln Y_{jt} = \alpha + \beta_1\, TreatmentGroup_j + \beta_2\, t + \beta_3 (TreatmentGroup_j\ x\ t) + \varepsilon_{jt} \quad (3)$$

where $TreatmentGroup_j$ is a dummy variable indicating whether municipality j is in the treatment group, $t$ is a time counter, $\alpha$ is the intercept (control group baseline), $\beta_1$, $\beta_2$, and $\beta_3$ are regression coefficients to be estimated, and $\varepsilon_{jt}$ is the stochastic disturbance term. If the coefficient of the interaction term, $\beta_3$, is significant, this would indicate a significant difference in the trend between the two groups.

Furthermore, a *placebo test* is frequently recommended in the literature (Goldfarb et al., 2022; Wang et al., 2024). In our case, a regression model is created using only the control group to check whether there is also a significant effect here by chance. To this end, the control group is randomly split in half into a pseudo-treatment group and a (remaining) control group; the DiD models (Equations 1 and 2) are then re-estimated using this split. If the corresponding treatment coefficients are nevertheless significant, this is interpreted as a placebo effect.

Since the case at hand involves only a single treatment unit (the city of Holzminden), a DiD analysis using a synthetic control unit is additionally conducted. A synthetic control unit (SCU) consists of the weighted average of the control group units (*donor pool*), with the weights iteratively optimized to best replicate the outcome of the treated unit during the pre-treatment period; this procedure is explicitly designed for research cases where the number of treatment units is substantially low. Combining DiD and SCU means that the SCU is used instead of the control group (Li et al., 2024). In our case, the SCU is constructed by minimizing the squared residuals between the weighted control unit values and the treatment city outcome, both transformed via natural logarithm, and given the constraint that the weightings sum up to one. The SCU's monthly overnight stay values, $Y_{SCU,t}$, equal thus:

$$Y_{SCU,t} = \sum_{j=1}^{J} \mu_j \, Y_{jt} \quad (4)$$

where $\mu_j$ is the weighting of municipality $j$.

Unlike the DiD models above, the equations of the models for the SCU data do not include unit fixed effects but instead a control group baseline and a treatment group deviation, with the latter being expected not to differ significantly from zero:

$$\ln Y_{jt} = \alpha + \beta \, TreatmentGroup_j + \lambda_t + \delta \, Sensoria_{jt} + \sum_{c=1}^{C} \beta_c \, V_{c_{jt}} + \varepsilon_{jt} \quad (5)$$

$$\ln Y_{jt} = \alpha + \beta \, TreatmentGroup_j + \lambda_t + \delta_1 \, Sensoria1_{jt} + \delta_2 \, Sensoria2_{jt} + \sum_{c=1}^{C} \beta_c \, V_{c_{jt}} + \varepsilon_{jt} \quad (6)$$

where $TreatmentGroup_j$ is a dummy variable indicating whether municipality $j$ is the treatment city (1) or not (0), $\alpha$ is the intercept (SCU baseline), and $\beta$ is the treatment group deviation.

The inference statistics for the coefficients of the DiD models including the control group are calculated using standard errors clustered at the level of the observational units (municipalities). The minimum accepted significance level is set at 90% ($p < 0.1$). The complete analysis was performed in *Python* v3.13.13 (Python Software Foundation, 2026) using the *diffindiff* package v2.5.4 (Wieland, 2026).

### 3.2 Estimating direct and indirect effects

In the second step, the additional overnight stays due to the museum opening are calculated first; we define this as the difference between the observed overnight stays in the municipality affected by the museum opening (Holzminden) and the expected overnight stays without the intervention, with the latter being calculated using the DiD equations (see section 3.1). These additional overnight stays are used to calculate gross turnover in the hospitality industry (hotels, restaurants), in retail and with respect to other services:

$$T_{b,Y} = \left(\sum_{t=1}^{M} Y_{Hol,t} - E_{Hol,t}\right) avex_b \quad (7)$$

where $T_{b,Y}$ is the additional gross turnover in industry $b$, $Y_{Hol,t}$ equals the observed overnight stays in the treatment city (Holzminden) in month $t$, $E_{Hol,t}$ is the expected overnight stays in the treatment city in month $t$, $M$ is the number of months in the intervention period, and $avex_b$ equals the average expenditure in industry $b$ per overnight stay. These expenditure values were obtained from several sources (see section 3.4). The industry definition is based on the NACE Rev. 2 classification of economic sectors (see Table 1).

*Table 1: Specification of regarded industries by economic sector classification.*

| Industry | WZ2008/NACE Rev. 2 classifications | Including |
|---|---|---|
| Hospitality | Section I = Division 55 Accommodation + Division 56 Food and beverage service activities | Hotels and similar accommodation, restaurants, bars, etc. |
| Retail sale | Division 47 Retail trade, except of motor vehicles and motorcycles – Group 47.9 Retail trade not in stores, stalls or markets | All physical retail stores, stalls and markets |
| Other services | Group 49.3 Other passenger land transport + Section R (Division 90-93) Arts, entertainment and recreation | Non train public transport (bus, taxi); Swimming baths, fitness facilities, libraries, museums, amusement parks etc. |

From the estimated gross turnover, the direct and indirect effects are derived. This calculation is based on the methodology used in the tourism impact studies by dwif (see, e.g., dwif, 2024a, 2024b; Federal Ministry for Economic Affairs and Energy, 2017). In the first step, the value added tax is deducted from the estimated gross turnover:

$$t_{b,Y} = T_{b,Y}\left(1 - \left(\frac{\theta_b}{100}\right)\right) \quad (8)$$

where $t_{b,Y}$ equals the net turnover attributable to the additional overnight stays, and $\theta_b$ is the (average) value added tax in industry $b$. The industry-specific value added tax $\theta_b$ was obtained from official statistics, namely the "Enterprises (EU), persons employed, turnover, production value and other business figures" table from the Federal Statistical Office of Germany (Destatis) (see Table 2). Note that the goods and services of individual sectors often consist of goods eligible for preferential VAT treatment (e.g., food) and goods not eligible for preferential VAT treatment (e.g., clothing); therefore, a weighted average of the preferential (7%) and the full VAT rate (19%) is usually calculated. The most up-to-date data used here is from 2021; these VAT rates essentially correspond to those in 2026.

*Table 2: Calculation of industry-specific average value added tax.*

| Industry | Taxable turnover 2021 [Million EUR] | VAT before input tax deduction 2021 [Million EUR] | $\theta_b$ = Average VAT [%] |
|---|---|---|---|
| Hospitality* | 70,888 | 7,484 | **10.56** |
| Retail sale** | 666,237 | 98,958 | **14.85** |
| Other services*** | 56,989 | 6,793 | **11.92** |

**NACE Rev. 2: Section I – Accommodation and Food Service Activities; **NACE Rev. 2: 47 Retail trade, except of motor vehicles and motorcycles, excluding 47.9 Retail trade not in stores, stalls or markets; *** NACE Rev. 2: 49.3 Other passenger land transport + Section R – Arts, entertainment and recreation. Sources: Destatis (2026a), own calculations.*

From the net turnover, in the second step, we calculate the production value:

$$PV_{b,Y} = t_{b,Y}\left(\frac{\psi_b}{100}\right) \quad (9)$$

where $PV_{b,Y}$ equals the production value in industry $b$ attributable to the additional overnight stays in the treatment city, and $\psi_b$ is the average share of production value in net turnover in industry $b$. The production value represents the value of sales of goods and services from own production to other economic entities, plus the value of the change in inventories of semi-finished and finished goods from own production and the value of self-constructed facilities. Since this value only refers to self-produced goods, it corresponds almost entirely to the net turnover in the hospitality industry, for example, while it is significantly lower in the case of retailing, as the value of purchased goods that are resold unprocessed is deducted from net sales (Destatis, 2023).

In the third step, the production value is split into direct effects (gross value added, i.e., the share of the production value that generates income, namely wages and entrepreneurial profit) and indirect effects (intermediate inputs):

$$DE_{b,Y} = PV_{b,Y}\left(\frac{\omega_b}{100}\right) \qquad (10)$$

$$IE_{b,Y} = PV_{b,Y} - DE_{b,Y} \qquad (11)$$

where $DE_{b,Y}$ and $IE_{b,Y}$ are the direct and indirect effects, respectively, with respect to industry $b$ based on the additional overnight stays in the treatment city, and $\omega_b$ is the average share of gross value added in the production value in industry $b$. The calculation of the relevant industry-specific coefficients, $\psi_b$ and $\omega_b$ is based on the industry average from official statistics, namely the "Enterprises (EU), persons employed, turnover, production value and other business figures: Germany, years, economic activities" table from Destatis (see Table 3).

*Table 3: Calculation of industry-specific share of production value and share of gross value added.*

| **Industry** | **Net turnover 2023 [Million EUR]** | **Production value 2023 [Million EUR]** | **$\psi_b$ = Share of production value from turnover 2023 [%]** | **Gross value added at factor cost 2023 [Million EUR]** | **$\omega_b$ = Gross value added ratio [%]** |
|---|---|---|---|---|---|
| Hospitality* | 127,171 | 122,985 | **96.71** | 62,693 | **50.98** |
| Retail sale** | 653,753 | 234,023 | **35.80** | 134,097 | **57.30** |
| Other services*** | 71,973 | 65,605 | **91.15** | 38,258 | **58.32** |

**NACE Rev. 2: Section I – Accommodation and Food Service Activities; **NACE Rev. 2: 47 Retail trade, except of motor vehicles and motorcycles, excluding 47.9 Retail trade not in stores, stalls or markets; *** NACE Rev. 2: 49.3 Other passenger land transport + Section R – Arts, entertainment and recreation. Sources: Destatis (2026b), own calculations.*

## 3.3 Definition of control group and study period

The treatment city is Holzminden, where the experience museum under investigation was opened. To test for a causal effect of this attraction on monthly overnight stays, a suitable control group unaffected by the opening is required. The difference-in-differences model explicitly accepts groups with different averages in the outcome and time effects independent of the treatment (Wooldridge, 2012); however, the control group should be meaningfully comparable to the treatment unit or group. In the first step, to narrow the control group down to areas that are comparable in cultural, and, above all, climatic, terms (which may directly influence tourism demand), the study area was limited to the (former) Regional Association of Southern Lower Saxony (*Regionalverband Südniedersachsen*). This (former) planning association includes the city of Holzminden as well as the entire counties of Northeim and Göttingen, with the latter being composed of the former counties of Göttingen and Osterode am Harz (Regionalverband Südniedersachsen, 2016).

*Figure 1: Flow diagram of control group definition (Own illustration using ChatGPT).*

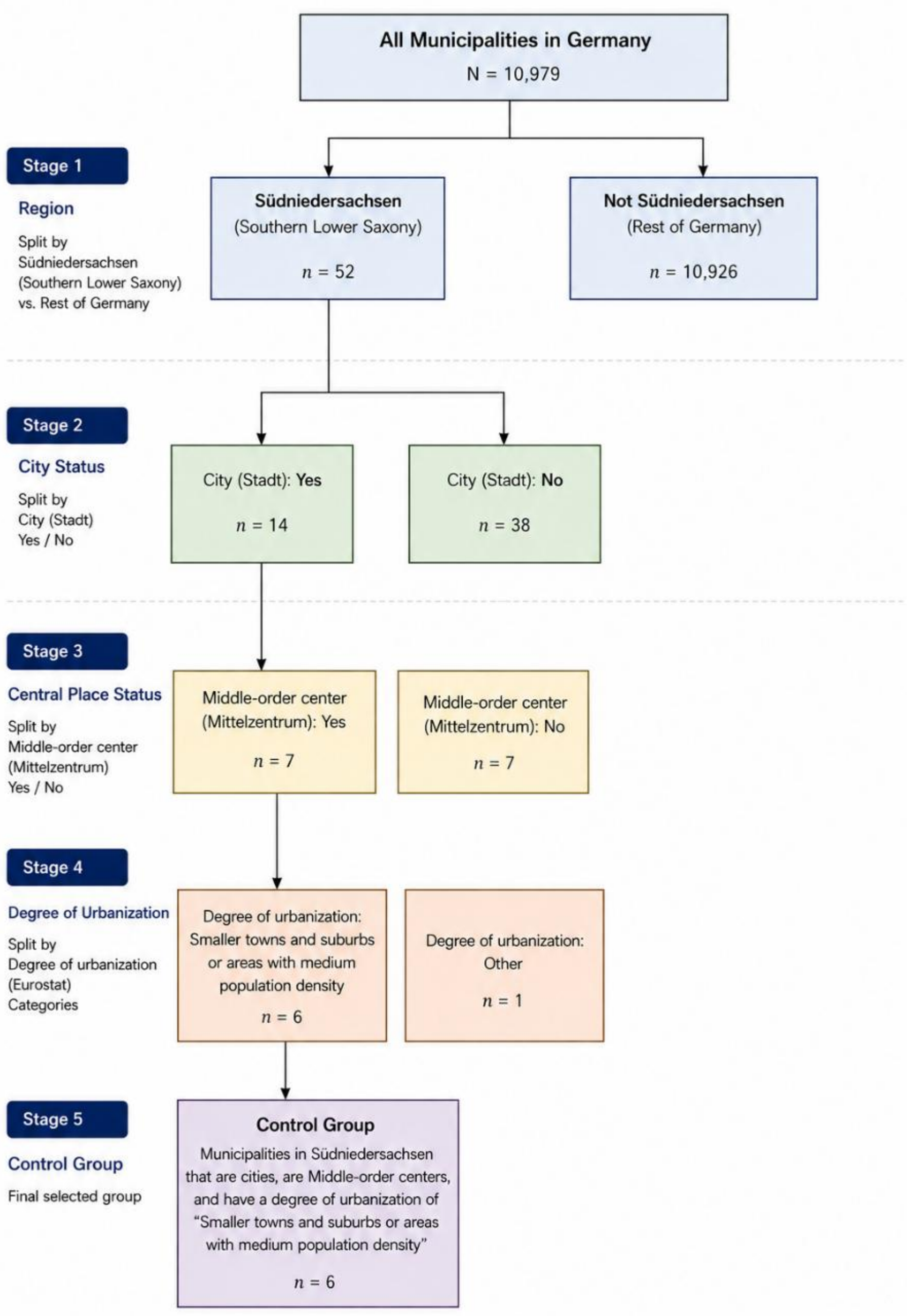

In the second step, the municipalities within this region were included based on their similarity to the treatment city in terms of their spatial-structural characteristics and their function within the spatial planning hierarchy. The treatment city of Holzminden is classified as a city ("Stadt"), which is defined by the German Federal Office for Building and Regional Planning (BBSR) as a municipality with at least 5,000 inhabitants or at least a basic central function within the spatial planning system (BBSR, 2025). It has the rank of a "middle-order center" ("Mittelzentrum") in the central place planning hierarchy of the *Lower Saxony State Spatial Planning Programme*. In terms of the Eurostat degree of urbanization, Holzminden has the classification "Smaller towns and suburbs or areas with medium population density" (BBSR, 2025; Ministry for Food, Agriculture and Consumer Protection of Lower Saxony, 2017). Based on the BBSR municipal classifications as of 2023 (BBSR, 2025), all municipalities classified differently were excluded from the group of South Lower Saxon municipalities, ultimately leaving a control group of six municipalities (see the flow diagram in Figure 1). These municipalities are: Bad Gandersheim, Northeim, Einbeck (Northeim county), Duderstadt, Hann. Münden, and Osterode am Harz (Göttingen county).

When defining the study period, care must be taken to select a sufficient and comparably long timeframe prior to the start of the treatment (September 2024). At the same time, the period must not extend so far back that potential travel restrictions associated with the coronavirus pandemic could induce a bias. As there were related travel restrictions in place in at least some countries until 2022, the start of the investigation period was set to January 2023. The end of the study period is also determined by the availability of statistical data on overnight stays. All statistics on monthly overnight stays that were available up to August 30, 2026, were included. In this case, that is the data up to and including June 2026. The study period thus includes 3.5 years (42 months) from January 2023 to June 2026, covering the first 22 months following the opening of *Sensoria* and 20 months prior to its opening, thereby resulting in a relatively balanced pre-treatment and treatment period. Given one treatment city and six control units, the number of observations in the DiD models thus equals 7*42 = 294, and 2*42 = 84 in the DiD models including the synthetic control unit.

### 3.4 Data collection

Monthly overnight stays per municipality and the number of bed days offered were obtained from the official tourism statistics (*Monatserhebung im Tourismus*). This is a full census in which all arrivals, overnight stays, beds offered, etc. are recorded in accommodation establishments with at least ten beds, covering the economic sectors (NACE Rev. 2) 55.1 (Hotels and similar accommodation), 55.2 (Holiday and other short-stay accommodation), and 55.3 (Camping grounds, recreational vehicle parks and trailer parks), as well as preventive and rehabilitation clinics and training centers (Destatis, 2025). The data for Lower Saxony municipalities are published by the Lower Saxony State Office for Statistics in tables K7360126 and K7361126 (Landesamt für Statistik Niedersachsen, 2026).

The data on events in the municipalities under investigation were collected via desktop research. For this purpose, the websites of the treatment city and the cities in the control group (see section 3.3) were systematically searched for all events occurring during the study period, occasionally utilizing officially published event calendars. All events lasting at least two days and taking place within January 2023 and June 2026 were recorded. Their durations were summarized as *event days* on a monthly basis. It is distinguished between four types of events for which four control variables were calculated: one each for small events (typically two to four days) and Christmas markets (typically three to five weeks), and one each for two multi-week events that took place in one municipality from the control group (*Landesgartenschau* and *Domfestspiele*, Bad Gandersheim). If there were multiple events in a municipality in the same month and the total number of event days would have been greater than the number of days in the respective month, the number of event days was set to the maximum number of days in the respective month (28-31).

Data on average overnight guest spending was not available for the treatment city, so secondary data sources from comparable municipalities in the same region were used. The data basis consists of two reports on the tourism impact in the city of Einbeck and Northeim county, respectively, which were prepared by dwif (2024a, 2024b), and the results of a visitor survey of cyclists on the Weser Cycle Route (Weserbergland Tourismus e.V., 2024), to which the treatment city belongs. All data is based on visitor surveys from 2023. Tourist spending is presented as an average EUR amount, broken down by guest type (overnight guests, daily visitors) and affected industry (hospitality, retail, other services). The arithmetic mean of the three values for overnight guests was calculated and used as an estimate for the average industry-specific expenditure per overnight stay ($avex_b$, see section 3.2).

# 4 Results and discussion

## 4.1 Descriptive statistics

Table 4 presents the descriptive statistics for the relevant indicators collected, including the outcome variable (overnight stays). The overnight stays and the event days from the four control variables regarding events are summed up per year, while sleeping accommodation days offered are expressed in terms of the arithmetic mean per month. With approximately 80,000 overnight stays per year, the treatment city falls below the arithmetic mean of the control group (2023: 112,186; 2024: 106,142; 2025: 105,587); however, this does not pose a limitation for the difference-in-differences analysis, as differing baselines between the groups are controlled for. Guest overnight stays in the base year of 2023 range from 52,345 (Northeim) to 190,433 (Bad Gandersheim), with the exceptionally high figure for the latter city most likely attributable to a major event held there (*Landesgartenschau*). The variation in the available sleeping accommodations is significantly lower, indicating widely differing occupancy rates among the municipalities.

*Table 4: Overnight stays, bed offered, event days in the municipalities investigated, 2023-2026.*

| **Municipality** | **Overnight stays (Yearly sum)** | | | | **Sleeping accommodation days offered (Monthly average)** | | | | **Event days (Yearly sum)** | | | |
|---|---|---|---|---|---|---|---|---|---|---|---|---|
| | **2023** | **2024** | **2025** | **2026*** | **2023** | **2024** | **2025** | **2026*** | **2023** | **2024** | **2025** | **2026*** |
| Bad Gandersheim | 190,433 | 173,712 | 173,691 | 82,889 | 35,214 | 35,218 | 36,121 | 33,563 | 199 | 76 | 82 | 17 |
| Duderstadt | 87,952 | 91,007 | 81,938 | 42,814 | 28,477 | 28,279 | 27,733 | 28,973 | 34 | 29 | 33 | 2 |
| Einbeck | 98,701 | 97,230 | 99,468 | 46,224 | 23,399 | 23,852 | 23,275 | 22,415 | 40 | 43 | 44 | 4 |
| Hann. Münden | 171,359 | 157,765 | 158,426 | 71,456 | 52,020 | 51,334 | 51,223 | 49,336 | 50 | 49 | 47 | 9 |
| **Holzminden** | **81,467** | **79,628** | **80,845** | **39,629** | **45,924** | **48,028** | **47,595** | **45,647** | **44** | **49** | **52** | **7** |
| Northeim | 52,345 | 53,089 | 50,915 | 23,417 | 24,033 | 24,076 | 20,311 | 20,285 | 60 | 57 | 59 | 2 |
| Osterode am Harz | 72,326 | 64,049 | 69,084 | 29,137 | 27,005 | 24,540 | 25,327 | 24,911 | 28 | 29 | 41 | 0 |

**January to June 2026. Sources: Landesamt für Statistik Niedersachsen, 2026; own data collection; own calculations.*

In total, 87 events with a duration of at least two days were recorded within the 3.5 years study period, resulting in 1,186 event days. The Christmas markets last an average of 30.1 days, while the small events (lasting 2 days or more) span an average of 3.4 days. The comparatively low figure for the first half of 2026 across all the municipalities studied is primarily attributable to the fact that Christmas markets account for a large proportion of the event days. Here, too, an extreme value for 2023 can be observed for the city of Bad Gandersheim within the control group, a figure likewise attributable to the fact that it served as the host town for the *Landesgartenschau* (185 days). Apart from that, the number of event days has remained relatively constant over the years, and the individual annual events also tend to take place in the same month. Apart from the exceptional year 2023, the treatment city is roughly at the level of the control group's arithmetic mean (2023: 68.5, 2024: 47.2, 2025: 51.0, first half of 2026: 5.7).

In table 4, the secondary data with respect to the average expenditure per overnight stay is presented. The variance in spending in the hospitality industry is relatively low; on average, 82.49 EUR (SD = 6.74 EUR) is assumed for accommodation and restaurants. For spending on retail and other services, the arithmetic means are 17.24 EUR (SD = 2.64 EUR) and 19.77 EUR (SD = 7.63 EUR), respectively.

*Table 5: Average expenditures [EUR] per overnight stay and industry.*

| **Industry** | **Tourists with overnight stays on the Weser Cycle Route 2023** | **Overnight guests in Einbeck 2023** | **Overnight guests in Northeim county 2023** | ***$avex_b$* = Arithmetic mean** | **Standard deviation** |
|---|---|---|---|---|---|
| Hospitality | 88.00 | 86.47 | 73.01 | **82.49** | 6.74 |
| Retail sale | n/a | 19.89 | 14.60 | **17.24** | 2.64 |
| Other services | 12.00 | 17.17 | 30.15 | **19.77** | 7.63 |

*Sources: dwif (2024a, 2024b), Weserbergland Tourismus e.V. (2024), own calculations.*

## 4.2 Difference-in-difference results

Table 6 shows the results (coefficients with standard errors and type I error probability, R-squared) for the four regression models, a DiD model with a control group and a DiD model including a synthetic control unit, each covering the entire study period and the treatment split into two time periods. The DiD models explain a high variance with $R^2$ = 0.930 and 0.931, respectively, which is also due to including fixed effects for both observational units and time points. In both models, no significant difference in the temporal trends between the intervention and control groups was found in the pre-period, meaning the parallel trend assumption is not violated based on this test. The placebo tests yielded negative results in both cases, indicating no artificial effect in the control group. The DiD models with the SCU both achieve a value of $R^2$ equal to 0.985. As expected, the parallel trends assumption is satisfied here, since the construction of the SCU is based on minimizing the deviation between the values of the weighted control group and the treatment city. A placebo test could not be conducted here, as there is only one control unit (the SCU).

*Table 6: Results of the difference-in-differences regression models.*

| | **DiD with control group** | | **DiD with synthetic control unit** | |
|---|---|---|---|---|
| **Dependent variables** | **Model 1** | **Model 2** | **Model 1** | **Model 2** |
| Treatment Sensoria | 0.043 (0.031) | -- | 0.042 (0.036) | -- |
| Treatment Sensoria first year | -- | 0.064** (0.027) | -- | 0.042 (0.042) |
| Treatment Sensoria after first year | -- | 0.018 (0.040) | -- | 0.043 (0.046) |
| *Control variables* | | | | |
| ln Bed days offered | 0.187 (0.256) | 0.189 (0.256) | -0.188 (0.206) | -0.189 (0.213) |
| Event days *Landesgartenschau* | -0.004* (0.002) | -0.004* (0.002) | -0.000*** (0.000) | -0.000*** (0.000) |
| Event days *Domfestspiele* | -0.009** (0.004) | -0.009** (0.004) | -0.000*** (0.000) | -0.000*** (0.000) |
| Event days Christmas markets | -0.015* (0.008) | -0.015* (0.008) | -0.034*** (0.010) | -0.034*** (0.011) |
| Event Days Small events | 0.006 (0.009) | 0.006 (0.009) | 0.005 (0.013) | 0.005 (0.013) |
| Control group baseline | -- | -- | 10.046*** (2.078) | 10.056*** (2.147) |
| Treatment group deviation | -- | -- | 0.098 (0.110) | 0.049 (0.057) |
| *Fixed effects* | | | | |
| Regions | Yes | Yes | No | No |
| Time points | Yes | Yes | Yes | Yes |
| $R^2$ | 0.930 | 0.931 | 0.985 | 0.985 |
| No. of observations | 294 | 294 | 84 | 84 |
| Parallel trends | Yes | Yes | Yes | Yes |
| Placebo effect | No | No | -- | -- |

** p < 0.1, ** p < 0.05, *** p < 0.01.*

Regarding the DiD models, in model 1, which included the entire opening time of the *Sensoria* museum as treatment, the difference-in-difference coefficient is positive but not significant. In model 2, where the opening of *Sensoria* is divided, a significant positive average treatment effect of $\delta \approx 0.064$ is found in the first year of opening ($p < 0.05$). This corresponds to an average increase in monthly overnight stays of $100*(\exp(\delta)\text{-}1) \approx 6.61\%$. However, no significant effect of the *Sensoria* opening was found for the period after the first year of opening; the corresponding coefficient is positive as well. The control variables are partially significant, with the event days variables not indicating positive impacts on overnight stays. However, this should not be interpreted to mean that the events have no (or even a negative) effect on tourism demand: it should be noted here that the local hospitality industry may increase its accommodation capacity for the duration of such events, which is already controlled out in the DiD models. Figure 2 shows the observed and expected time series of overnight stays in the treatment and control groups based on DiD model 2.

*Figure 2: Overnight stays in the treatment city and control group, observed and estimated (DiD model 2).*

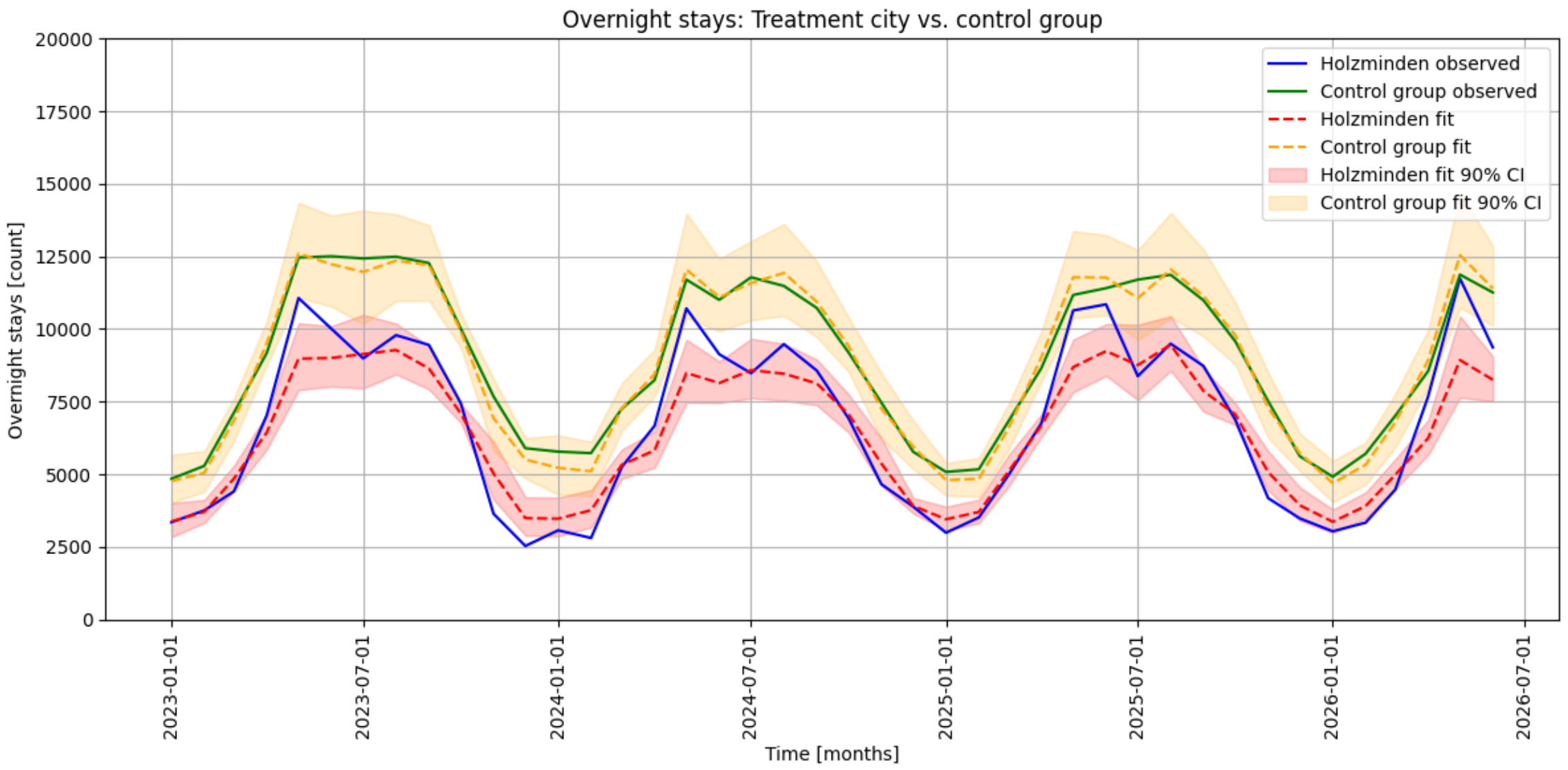

The synthetic control unit covering the pre-treatment period shows an adequate fit to the empirical overnight stays of the treatment city (R-squared: 0.925, MAPE: 1.19 %; see also Figure 3). The percentage contributions, i.e., empirically determined weightings, of the analysis units from the donor pool are presented in Table 7. The DiD models including the SCU yield different results than the DiD models, which is entirely plausible, given that the DiD model implicitly weighs all units in the control group equally, whereas the DiD models including the SCU apply specific weights. In addition, it must be taken into account that combining the six municipalities of the control group into one SCU significantly reduces the number of observations, which may lead to substantially higher standard errors. In the first SCU model, the average treatment effect is nominally about the same as in the first DiD model and is likewise not significant. In the second SCU model, both treatment coefficients are positive (as in the second DiD model) but not significant.

*Table 7: Weightings of the donor pool municipalities.*

| **Municipality** | **Weigthing [%]** |
|---|---|
| Bad Gandersheim | 0.00 |
| Duderstadt | 0.00 |
| Einbeck | 20.22 |
| Hann. Münden | 14.00 |
| Northeim | 7.39 |
| Osterode am Harz | 58.40 |

*Figure 3: Overnight stays in the treatment city and the synthetical control unit during the pre-treatment period*

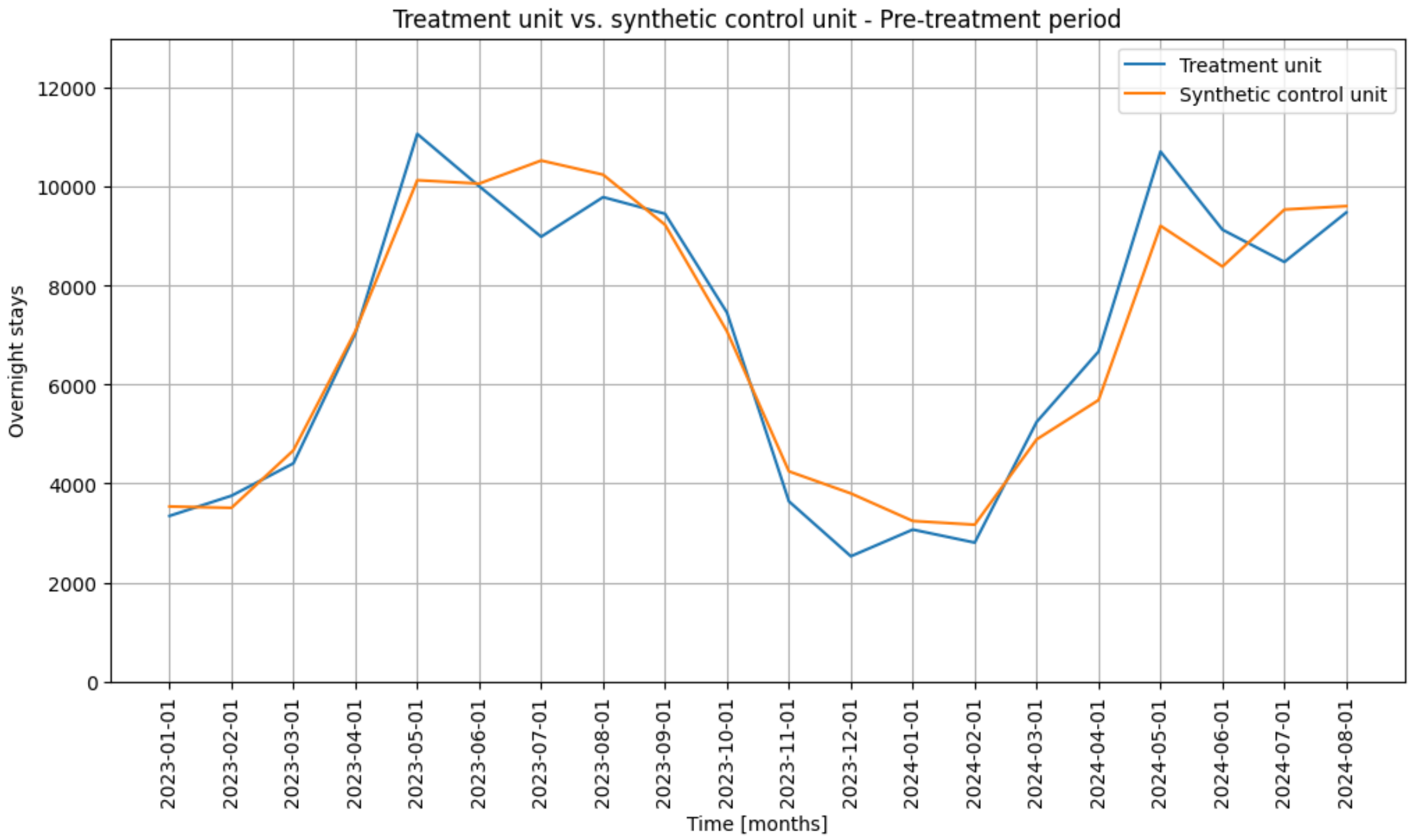

Although the model results are mixed, they suggest a positive treatment effect at least for the first year following the opening of *Sensoria*. This effect is significant in the DiD model but not in the synthetic control model, which is presumably due to the substantially smaller number of observations. The consistent direction points to a positive impact, although this is statistically uncertain in the second model. Consistent with this, the coefficients for the second period are positive, just like those for the entire period in the respective first models, although all of them are subject to excessive statistical uncertainty.

The numeric results cannot be directly compared with the findings of other studies on the impact of tourist attractions on tourist demand, as the attractions examined are of a very different nature. However, previous studies regarding built tourism attractions with a cultural focus – particularly museums – conducted using various methodological approaches show that such facilities increase local tourism demand: For example, using an SCU approach, Eckhardt et al. (2026) found an uplift in overnight stays in Hamburg resulting from the opening of the *Elbphilharmonie*. Although the scale of that facility cannot be compared to the case of the experience museum examined here, both sets of results point in the same direction. In terms of the size of the host city and the focus and visitor numbers of the facility studied, the study by Llop and Arauzo-Carod (2012) on the *Gaudí Centre* in Reus, Catalonia, is the most comparable to the case at hand; they likewise identify additional tourism demand generated by that museum. The fact that an effect can be demonstrated for the first year of operation but not for the subsequent period could be attributed to the simple lack of sufficient data for the second period (only 10 months) to assess any long-term effect. However, it is also possible that such an effect manifests only during the opening year because visitor numbers are exceptionally high due to the novelty, whereas that volume is not reached in subsequent years. For instance, the study by Plaza (2006) indicates that visitor numbers at the *Guggenheim Museum* in Bilbao were significantly higher in the first year than in the years that followed.

## 4.3 Direct and indirect effects

The calculation of direct and indirect effects is based on the first year of *Sensoria*'s opening, as a reliable and significant positive effect could only be estimated for this period. Figure 4 shows the observed overnight stays in the treatment city and the estimated counterfactual for the treatment period in DiD model 2. From September 2024 up to and including August 2025, a cumulative 4,964 additional overnight stays can be attributed to the new tourist attraction, which corresponds to an average of approximately 414 additional overnight stays per month.

*Figure 4: Treatment city observed and estimated counterfactual (DiD model 2).*

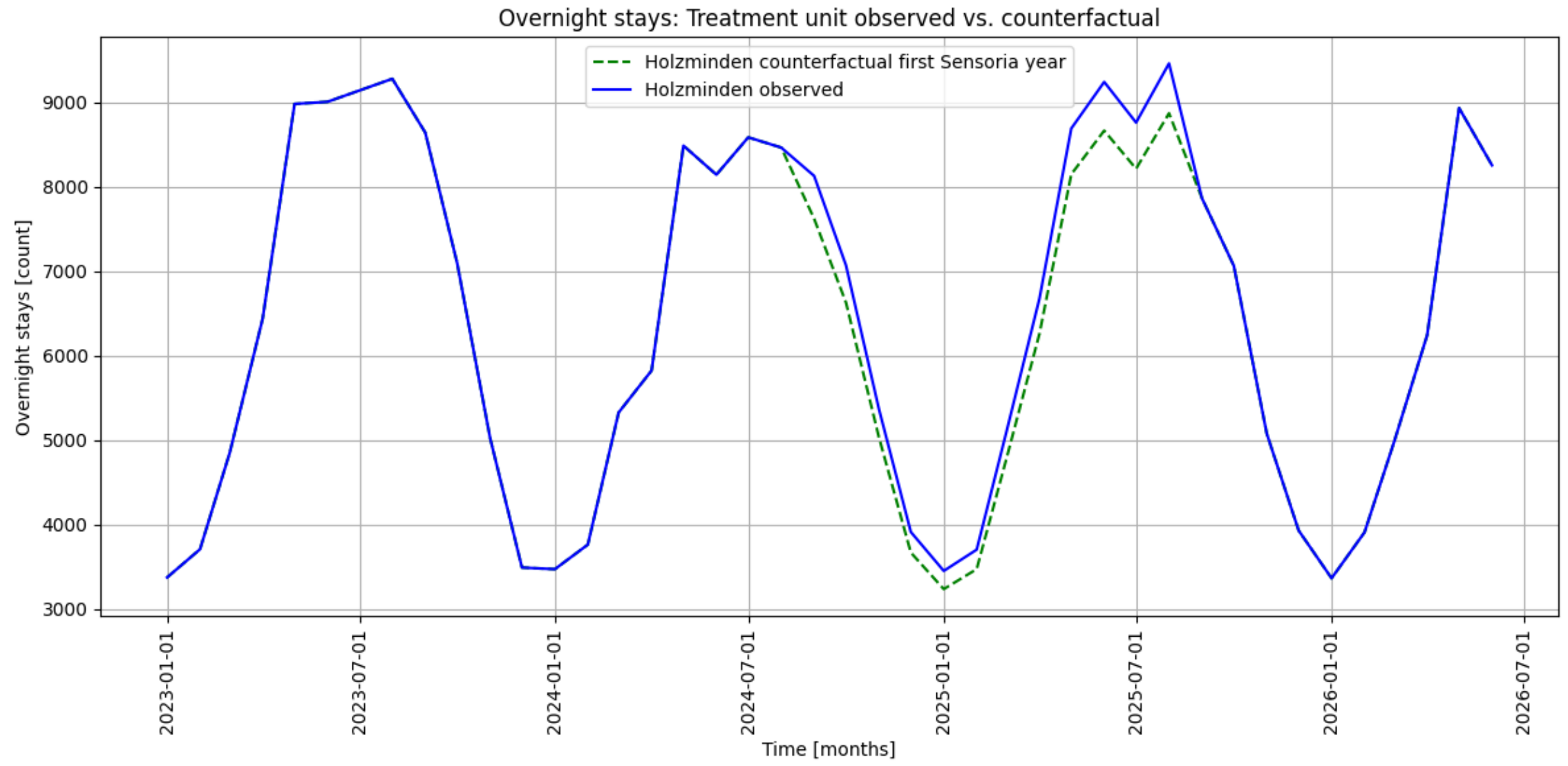

Based on the assumed on-site spending of these additional overnight guests, this results in a gross turnover of approximately EUR 0.6 million, the majority of which (EUR 0.41 million) is in the hospitality industry (see Table 8). When the direct and indirect effects are calculated from these gross turnovers, the direct effects (gross value added) and indirect effects (intermediate inputs) amount to approximately 0.24 and 0.22 million EUR, respectively, for the first year of *Sensoria*'s opening. The majority of this (approximately 75-80 %) is attributable to the hospitality industry (hotels, restaurants), which is due to the fact that 1) the majority of the assumed tourist expenditure is attributable to this economic sector (particularly due to accommodation) and 2) the ratio of production value to gross turnover is particularly high here (see Table 3 in the section 3.2).

*Table 8: Gross turnover and direct and indirect effects (based on DiD model 2).*

| Industry | Gross turnover [€] | Direct effects [€] | Indirect effects [€] |
|---|---|---|---|
| Hospitality * | 409,503 | 180,575 | 173,634 |
| Retail sale ** | 85,584 | 14,948 | 11,141 |
| Other services *** | 98,143 | 45,952 | 32,842 |
| **Total** | **593,230** | **241,475** | **217,617** |

**NACE Rev. 2: Section I – Accommodation and Food Service Activities; **NACE Rev. 2: 47 Retail trade, except of motor vehicles and motorcycles, excluding 47.9 Retail trade not in stores, stalls or markets; *** NACE Rev. 2: 49.3 Other passenger land transport + Section R – Arts, entertainment and recreation. Sources: own calculations.*

# 5 Conclusions and Limitations

Assessing the economic impact of tourist attractions is highly relevant, particularly in the case of public investments. Such assessments often rely on demand-side economic analyses based on visitor surveys. As has been demonstrated, this approach has significant limitations – notably that it does not constitute a causal analysis and that surveys can be subject to substantial bias. The present study therefore aimed to develop a methodological workflow that combines causal analysis techniques with the estimation of the direct and indirect effects of tourism facilities. Causal analysis was conducted using a difference-in-differences (DiD) model and a DiD model with a synthetic control unit. The case study focused on the opening of the *Sensoria* experience museum in Holzminden, Germany, examining the impact on tourist overnight stays. Regarding the case study, it can be concluded that a positive effect is likely. At least for the first year of operation, this effect can be empirically demonstrated and is estimated at approximately 5,000 additional overnight stays, or a contribution of roughly 0.5 million EUR.

With respect to the methodology, we may conclude that it is possible to combine both methods: in the first step, causal-analytical models were constructed, and the additional tourist overnight stays – which are very likely to be attributed to the new tourist attraction – were derived from the counterfactual. The workflow developed here offers a decisive advantage: by accounting for 1) a control group and 2) control variables that can influence tourism demand, it provides stronger causal evidence regarding the impact of the tourism facility. A conventional assessment of the impact of *Sensoria*'s opening would, in the simplest scenario, have been based on the number of the facility's visitors, categorizing them into day visitors and overnight guests; a more comprehensive approach would have involved a visitor survey to gather data on the nature of their visits and daily spending, while also attempting to identify those visitors whose presence in the treatment city was due solely (or primarily) to *Sensoria*. Another specific advantage of this approach is that – apart from any very minor errors in the official tourism statistics that may exist – there is no bias in the outcome (monthly overnight stays). A visitor survey could only guarantee this if it were sufficiently representative; this is not impossible, but appears difficult, given that one cannot assume all visitors possess the same amount of time and the same level of motivation to participate in a survey.

However, this novel combination of two methods entails a variety of challenges: these include, among others, the construction of an appropriate control group, the meaningful definition of the study period, the inclusion of suitable control variables, and the model specification itself. Regarding the last point, it should be noted that two modeling approaches were used – one employing the full control group and the other a synthetic control unit – whose results point in the same direction but differ numerically, including in terms of effect size. Both econometric analysis approaches can be equally well justified; it therefore makes sense to carry out both forms of analysis and compare their results.

The current study faces some limitations with respect to the specific case. *First*, the impact of the opening of the *Sensoria* museum was assessed using the official monthly tourism statistics, which do not include overnight stays in small or private establishments. Furthermore, day visitors are also not considered in this analysis, as their number is unknown. This has no impact on the econometric analysis, since the outcome variable is always operationalized identically; however, the failure to consider day visitors and some overnight visitors means that the direct and indirect effects in the second step of the analysis will in any case be underestimated. *Second*, the validity of the result is limited by the study period, as it only covers the first 22 months after the opening of the *Sensoria* museum. While it is possible to make well-founded statements about the first year of operation, this is not possible for the period beyond, which only covers four months. *Third*, a potential problem arises here that has also been criticized in relation to other studies: the estimation of direct and indirect effects is based on the results of surveys that were not conducted by the researchers themselves, making it impossible to verify whether they are representative or to determine the procedure used to select the participants.

Considering these limitations, it is therefore advisable to conduct a more comprehensive study after a specific period has elapsed (spanning several years; see, e.g., Plaza, 2006). This study should involve repeating the econometric analysis employed here, supplemented by a representative visitor survey designed to directly capture data from *Sensoria*'s visitors. This survey should record, among other things, on-site expenditure, reasons for visiting, and visitation modalities; a link between overnight stays and visitor numbers may be established by comparing the additional overnight stays with the overnight guests identified in the survey. The workflow presented here should also be applied to other case studies and critically evaluated.